\documentclass[prb,preprint,groupedaddress,endfloats]{revtex4}
\usepackage{bm,graphicx,graphics,amsmath,amssymb,bm,epsfig,color}
\usepackage{euscript,tabularx}
\usepackage{textcomp}

\begin{document}
\bibliographystyle{apsrev}
\preprint{\begin{minipage}{8cm} Number: \hrulefill
   \end{minipage}}

 
 \title{Spectroscopy of the parametric magnons excited by 4-wave process}

\author{G. de Loubens}
\affiliation{Service de Physique de l'{\'E}tat Condens{\'e}, CEA Orme des
  Merisiers, F-91191 Gif-Sur-Yvette}
\author{V. V. Naletov}
\affiliation{Service de Physique de l'{\'E}tat Condens{\'e}, CEA Orme des
  Merisiers, F-91191 Gif-Sur-Yvette}
\altaffiliation[Also at ]{Physics Department, Kazan State University, Kazan 420008 Russia}
\author{V. Charbois}
\affiliation{Service de Physique de l'{\'E}tat Condens{\'e}, CEA Orme des
  Merisiers, F-91191 Gif-Sur-Yvette}
\author{O. Klein}
\affiliation{Service de Physique de l'{\'E}tat Condens{\'e}, CEA Orme des
  Merisiers, F-91191 Gif-Sur-Yvette}
\thanks{Corresponding author} 
\email{oklein@cea.fr}
\author{V. S. Tiberkevich}
\affiliation{Department of Physics, Oakland University, Michigan US-48309}
\author{A. N. Slavin}
\affiliation{Department of Physics, Oakland University, Michigan US-48309}

\date{\today}

\begin{abstract}
  Using a Magnetic Resonace Force Microscope, we have performed ferromagnetic resonance (FMR) spectroscopy on parametric magnons created by 4-wave process. This is achieved by measuring the differential response to a small source modulation superimposed to a constant excitation power that drives the dynamics in the saturation regime of the transverse component. By sweeping the applied field, we observe abrupt readjustement of the total number of magnons each time the excitation coincides with a parametric mode. This gives rise to ultra-narrow peaks whose linewith is lower than $5~10^{-6}$ of the applied field.  
\end{abstract}

\maketitle

The detailed understanding of the non-linear (NL) regime of the
magnetization dynamics is important both from a fundamental point of
view \cite{slavin:06} but also for applications in spintronic devices
\cite{prinz:99}. Interest resides in the exact nature of the
parametric modes that are excited above the supercriciality threshold.
Recent experiments performed on a yttrium iron garnet (YIG) film have
shown that these parametric magnons can form a Bose-Einstein
condensate \cite{demokritov:06} under high power pumping. It was also
demonstrated that their energy decay rate to the lattice is
substantially diminished compared to the long wavelength modes
\cite{loubens:05}, which are usually studied by FMR In this paper, it
will be shown that it is also possible to measure the FMR spectrum of
these parameric magnons despite the fact that these high $k$-vector
spin-waves do not overlap with the homogeneous microwave field. This
is achieved by saturating the dynamics of the mode that couples to the
microwave. Any additional power that is injected in the spin system
flows directly towards the excitation of these parametric magnons.
Since each parametric mode has a different feedback influence on the
saturation, it is possible, by sweeping an external parameter, to
detect the point where the transfer from one parametric mode to
another takes place. But the induced effects are small and detecting
them requires both a sensitive and precise measurement setup.

\begin{figure}
\includegraphics[width=12cm]{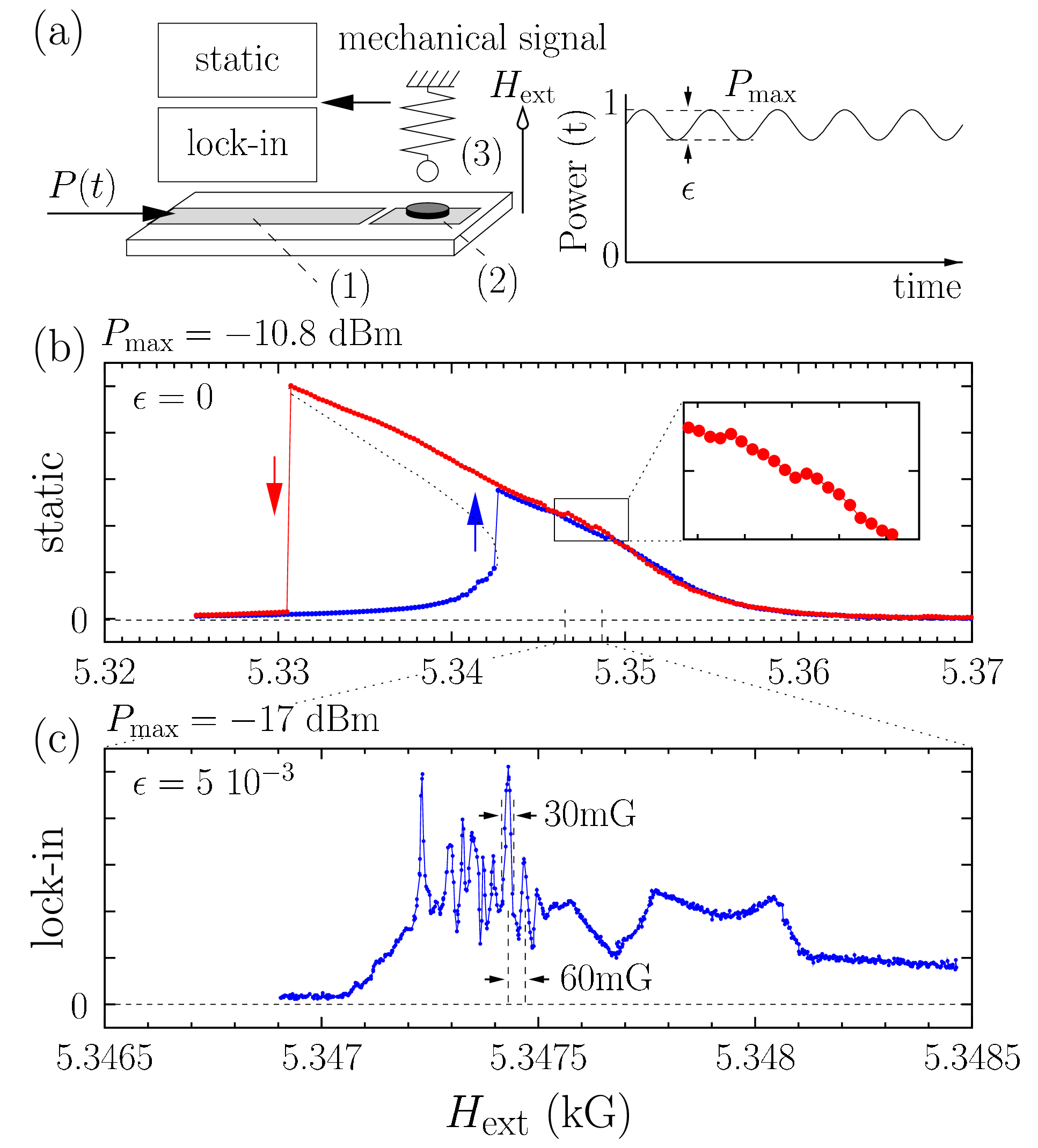}
\caption{(a) Schematic of the microstrip circuit (1). The FMR spectrum
  of a micron-size YIG disk (2) is detected by Magnetic Resonance
  Force Microscopy (3). On the right is shown $P(t)$, the waveform of
  the power injected in the microwave circuit (1). The amplitude of
  the sinusoidal modulation is $\epsilon P_\text{max} /2$. (b) Static
  deflection of the cantilever as a function of $H_\text{ext}$. The
  trace is the lineshape of the uniform mode in the foldover regime.
  (c) Vibration amplitude of the cantilever measured by a lock-in
  synchronous with source modulation. The differential
  characteristics reveals the parametric excitation spectrum.}
\label{fig1}
\end{figure}

Our detection of the dynamical response uses a mechanical-FMR setup
working at room temperature \cite{zhang:96,charbois:02}. This scheme
is inspired by local probe techniques. The static component of the
disk magnetization, $M_z= M_s \cos \theta$, is coupled through the dipolar
interaction to a magnetic probe attached at the end of a soft
cantilever, whose deflection is monitored. Exciting the sample at a
fixed frequency, spectroscopy is achieved by recording the cantilever
motion as a function of the perpendicular dc applied field,
$H_{\text{ext}}$, which is produced by a NMR electromagnet. The
variation of force on the cantilever is proportional to the total
number of magnons excited at resonance, $N_t$, with $N_t \equiv
(M_s-M_z)/(\gamma \hbar)$.  Further details about the mechanical-FMR experiment
can be found in Ref.\cite{naletov:03}.

Driving the dynamics of ferromagnets in the NL regime requires to cant
the magnetization, $M_s$, away from its equilibrium axis by an angle
$\theta$ exceeding a couple of degrees. Only powerful HF source can produce
such large excitation amplitude \cite{an:04}. Improvement of the
efficiency can be gained with microstrip cavities, where the HF energy
is concentrated in smaller volume. A schematic of our setup is shown
in Fig.\ref{fig1}(a). It comprises a $0.5$mm wide Au stripeline
fabricated by U.V. lithography and deposited on a $0.5$mm thick
alumina substrate whose bottom layer is a conducting ground plane. An
impedance matched cavity (half-wavelength) is created by etching a
32$\mu$m gap across the stripe.  The HF source is tuned at the cavity
frequency ($\omega_s / 2 \pi = 10.47$GHz) set by the length ($5$mm) of the
isolated segment. The sample is a $t=4.75 \mu$m thick YIG single
crystal.  The YIG film is ion milled into a disk of diameter $\phi =160
\mu$m and placed at the center of the cavity.  The homogeneous external
static field $H_{\text{ext}}$, well above the saturation field, is
applied parallel to the disk axis.  

This work concentrates on the 4-magnons coupling term in the equation
of motion of the magnetization. It becomes the dominant term once the
premature saturation regime of the microwave susceptibility is
achieved \cite{suhl:57}. From previous experiments on the same sample
\cite{loubens:05}, we have established that the saturation threshold
occurs at excitation power above -35dBm for our setup.
Fig.\ref{fig1}(b) shows the static deflection of the cantilever as a
function of $H_\text{ext}$ when $P_\text{max} =-10.8$ dBm. The trace
corresponds to the lineshape of the uniform mode \cite{charbois:jap}
in the foldover regime.  Hysteretic behaviors are a standard signature
of NL effects, where the eigen-frequency of the resonance depends on
the amplitude of the excitation \cite{lvov}. Foldover effects in
magnetic materials have been explained by Anderson and Suhl in 1955
\cite{anderson:55}. The features of interest are the tiny steps
observed in the wings of the resonance (see insert of
Fig.\ref{fig1}(b)).  These jumps are not random.  They are
reproducible and occur regularly in field.

A more detailed picture can be obtained by measuring the differential
response. We detect here the response to a source modulated excitation
around the saturation regime. The power level follows the time
dependence $P(t) = P_\text{max} \left \{ 1 + \frac{\epsilon}{2} (\sin (\omega_c
  t)-1) \right \}$, where the modulation frequency is set at the
resonance frequency of the cantilever. This frequency is much lower
than all the damping rates of the spin system. In this case, the
differential part of the mechanical signal is amplified by $Q=4500$,
the quality factor of the mechanical resonator.  Fig.\ref{fig1}(c) is
the pattern recorded by a lock-in when the modulation amplitude is
0.5\% ($\epsilon=5~10^{-3}$) of the maximum power, $P_\text{max} =-17$ dBm.
It corresponds to a source modulation of less than 50nW in amplitude.
The spectrum shown here is a zoom on a 2G sweep of $H_{\text{ext}}$ in
the reversible region of the static signal. We observe about ten
ultra-narrow lines (width 30mG) regularly spaced every 60mG. Such
sharp peaks resemble the parametric excitations observed by Jantz and
Schneider in 1975 in the subsidiary absorption of YIG films
\cite{jantz:75} (3-wave process).  To the best of our knowledge, this
is, however, the first time that they are observed at resonance.

\begin{figure}
\includegraphics[width=12cm]{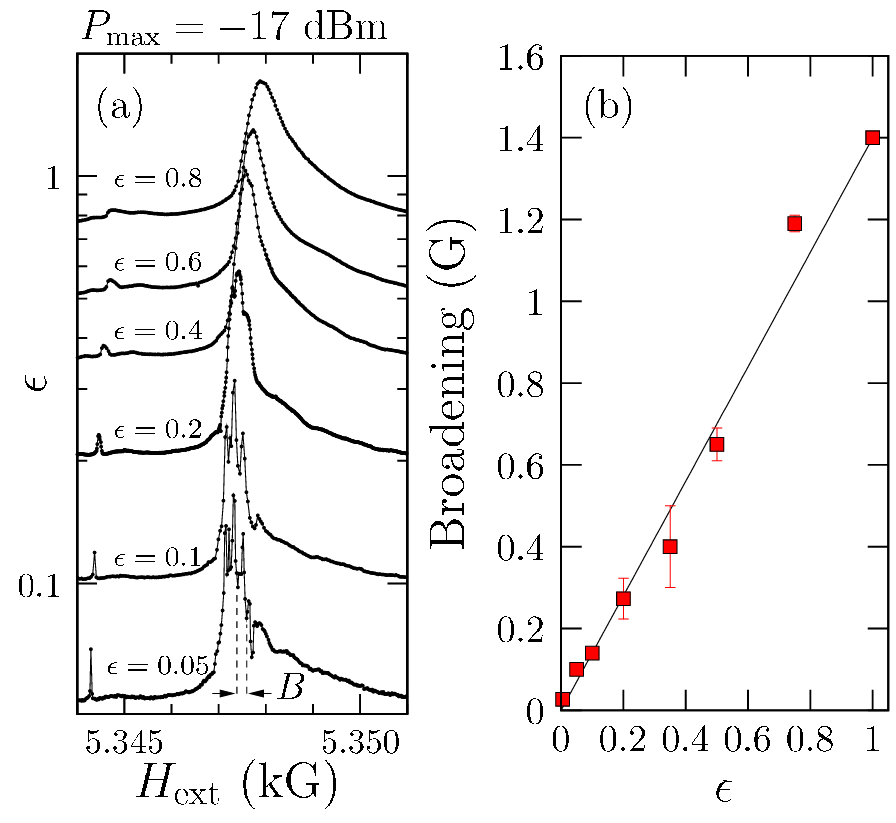}
\caption{(a) Differential spectrum as a function of the modulation depth
  $\epsilon$ (constant $P_\text{max}=-17$dBm). $B$ is the broadening of the parametric
  peaks. (b) $B$ as a function of $\epsilon$.}
\label{fig2}
\end{figure}

The striking characteristics of these peaks is their narrowness. This
30mG broadening is much smaller than the FMR linewidth $\Delta H=1.2$G
measured in the linear regime on the same sample \cite{klein:03}.
Further insight can be obtained by looking at the shape of the signal
(see Fig.\ref{fig2}a) for different modulation depth, $\epsilon$, at
constant maximum power $P_\text{max}$.  Fig.\ref{fig2}b shows that the
broadening $B$ of the parametric peaks is clearly proportional to $\epsilon$.
Foldover effects establish a coupling between the HF power and the
resonance frequency. Thus source modulation in the NL regime produces
also a frequency modulation. But in FMR, there is no difference
between frequency modulation and a modulation of the applied field,
since both are coupled by the gyromagnetic ratio. Fig.\ref{fig1}c can
thus been seen as the differential characteristics of Fig.\ref{fig1}b,
where the depth of modulation is proportional to $\epsilon$.  At very low $\epsilon$
eventually the intrinsic broadening should be seen. Although a NMR
electromagnet with a $10^{-6}$ precision in $H_{\text{ext}}$ has been
used, the fact that $B$ extrapolates linearly down to the smallest
value of $\epsilon$ reached by our setup, suggests that the intrinsic part of
the broadening has not been measured here.  

\begin{figure}
\includegraphics[width=12cm]{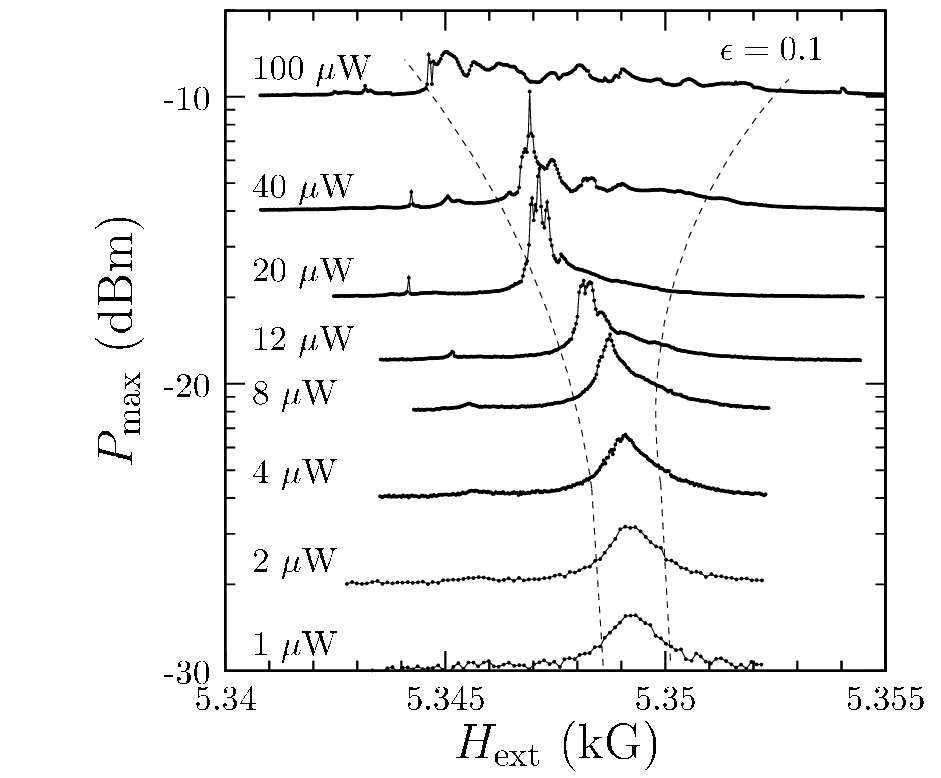}
\caption{Differential spectrum as a function of $P_\text{max}$ (constant
  $\epsilon=0.1$).}
\label{fig3}
\end{figure}

For completeness, we show in Fig.\ref{fig3} how the instability region
evolves when the dc bias $P_\text{max}$ is increased at constant $\epsilon$.
We observe an increase of the applied field window, where parametric
peaks occur. The size of the window follows monotically the amplitude
of the dc power \textit{i.e} how far we depart from the
supercriticality threshold.

In the following, we propose an analytical framework to account for
the main features observed experimentally. Saturation phenomena can be
caused by a multitude of nonlinear processes: nonlinear dissipation,
nonlinear phase mechanism, nonlinear energy feedback effect. In many
cases several of these mechanisms, as well as nonlinear frequency
shift, need to be accounted simultaneously for the correct description
of the phenomenon. The most probable limiting mechanism in the case of
perpendicular pumping is the nonlinear energy feedback.  This
mechanism is described by a 4-magnons interaction
$(\xi_kb_0^2b_k^*b_{-k}^* + {\rm c.c})$ in the Hamiltonian of the
system. It represents second-order parametric excitation of the pair
of plane spin waves ($b_k$, $b_{-k}$) by the uniform precession
($b_0$). Suhl showed in 1957, the equations of motion for the
amplitudes $b_k$ of the plane spin waves have the form:
\begin{equation}\label{bk}
  \dot{b_k}  = -i \omega_k b_k -i \xi_k {b_0}^2 {b_{-k}}^\ast - \eta_k b_k,
\end{equation}
where $\eta_k$ is its \emph{energy} decay to the lattice (the index must
allow for $k$-dependent relaxation rates \cite{loubens:05,hoeppe:05}).
The equation for the amplitude $b_0$ of the uniform precession is
given by
\begin{equation}\label{b0}
  \dot{b_0} = -i\omega_0 b_0 -i\sum_k \xi_k b_k b_{-k} b_0^\ast -\eta_0 b_0 + \gamma h e^{-i\omega_s t}  \ .
\end{equation}
where the last term is the excitation field $h=\sqrt{P}$. The
term $\sum_k\dots$ describes nonlinear feedback effect of all the
parametric waves on the uniform precession. This term gives effective
nonlinear dissipation for $b_0$. In the following, we will be looking
at harmonics solution where all the spin-waves precess
synchronously: $\dot{b_k}=-i \omega_s b_k$.

\begin{figure}
\includegraphics[width=12cm]{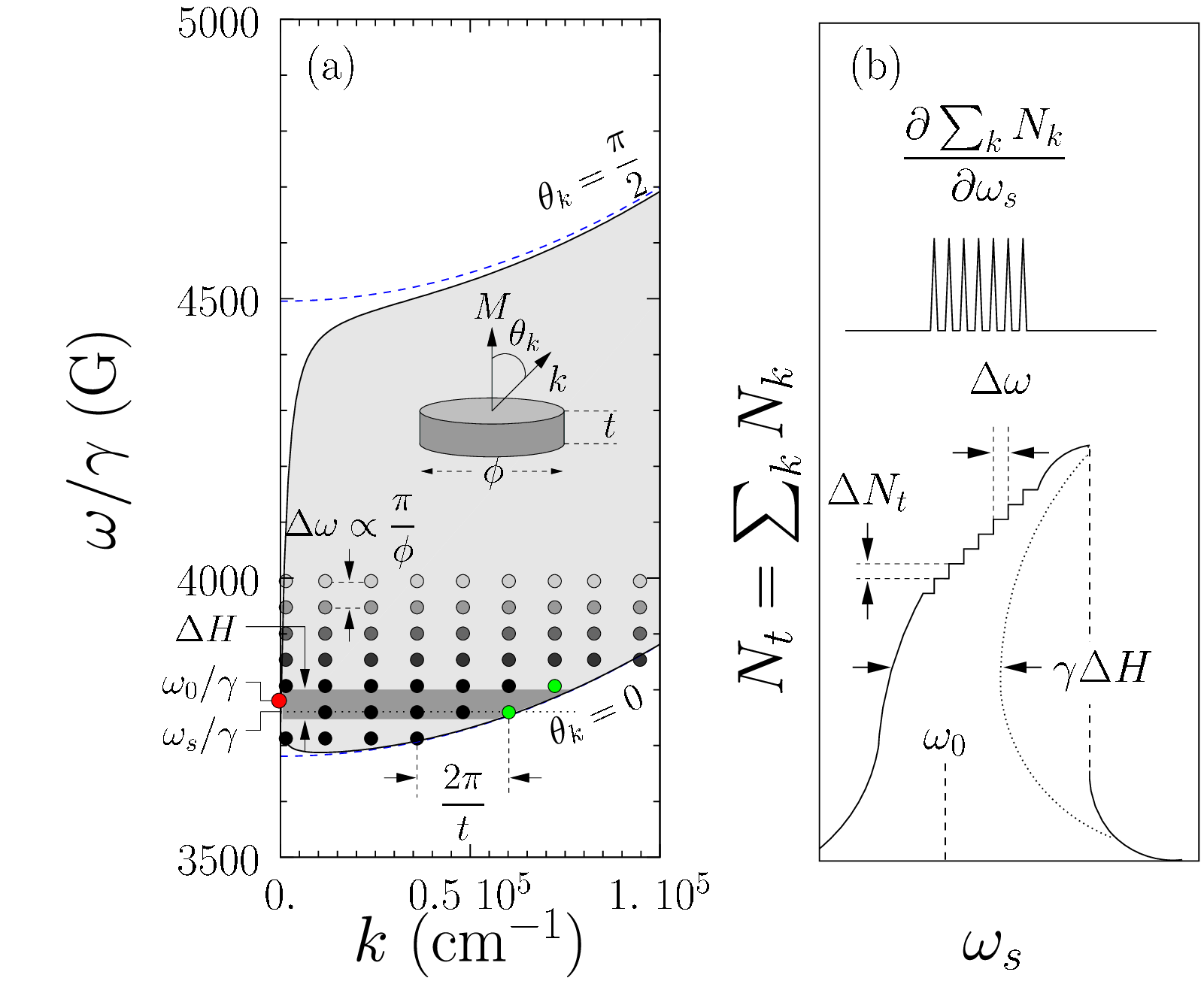}
\caption{(a) Magnon manifold of our perpendicularly magnetized YIG
  disk with a schematic of the mode discretisation induced by the
  confinement in a micron-size sample. Four-waves process couples the
  uniform mode (red) to the parametric modes at $\theta_k=0$ (green) (b)
  Schematic of the distorsion induced by the 4-waves NL process on the
  static spectroscopy of $M_z = \gamma \hbar \sum_k N_k$ vs. $\omega_s$. The staircase
  behavior reveals the parametric spectrum in the differential
  characteristics.}
\label{fig4}
\end{figure}

As one can see from Eq.~(\ref{bk}), an instability (exponential growth of
the population) occurs if the effective damping of at least one parametric
mode becomes negative:
\begin{equation}
  |b_0|^4 > \frac{(\omega_k - \omega_s)^2 + \eta_k^2}{\xi_k^2} \ .
\end{equation}
Growth of $b_k$ will create effective damping for the unifrom mode $b_0$
(through the $\sum_k\dots$ term in Eq.~(\ref{b0})), which will reduce $b_0$ to
its threshold value:
\begin{equation}\label{Bth}
	|b_0|^2 = N_0 \equiv \min_k\frac{\sqrt{(\omega_k-\omega_s)^2 + \eta_k^2}}{\xi_k} \,,
\end{equation}
Determination of the mode that will grow unstable depends on the
coupling $\xi_{k}$, which is maximum $\xi_{k} / \gamma = 2 \pi M_s (\approx 900$G) for
$k$ propagating parallel to the magnetization direction
\cite{suhl:57}. As shown on the magnon manifold drawn in
Fig.\ref{fig4}, these longitudinal SW have a wave-vector $k_\text{max}
=\sqrt{N_\perp4\pi M_s^2/(2A)}\approx 6.3~ 10^4 \text{cm}^{-1}$, where $A$ is the
exchange constant and $N_\perp$ is the transverse depolarization factor.
For our disk, it corresponds to a standing spin-wave confined across
the thickness with about 5 nodes along it. In this model, the
threshold power is given by the analytical formula, $h_c ^2 = \eta_k \Delta
H^2/ \xi_k$ and the calculated value $h_c =5$~mOe agrees with the the
onset of saturation measured experimentally \cite{loubens:05}.

Sweeping the bias magnetic field $H_\text{ext}$ (or $\omega_s$) changes the
nature of the parametric mode that grows unstable, as the minumum in
Eq.~(\ref{Bth}) occurs for different value of $k$.  Finite
size-effects introduce a discretisation in $k$-space of all the SW.
The frequency separation $\Delta\omega$ between two nearby normal mode is about
$\Delta \omega = 4 \gamma A k \Delta k /M_s$, where $\Delta k = \pi/ \phi$ corresponds to the
quantification of the $k$-vector along the largest sample dimension.
The parametric modes have thus a principal wavector $k_\text{max} \| M_s$,
with a small additional component $\Delta k$ in the disk plane. We find
numerically $\Delta \omega/ \gamma = 0.12$Oe, a predicted separation twice as large
as what is observed experimentally in Fig.\ref{fig1}(c). The value of
$k_\text{max}$ however depends strongly on the angle of propagation of
the degenerate magnons.  A deviation of about 7\textdegree{} of their propagation
direction compare to the normal of the disk will be enough to explain
the discrepancy.

The amplitude of the steps can also be estimated analytically with an
approximated model.  Assuming that only one pair of parametrically
coupled waves dominates all the others, then its amplitude can be
infered from Eq.~(\ref{b0}). If the mode $b_k$ is excited, then its
amplitude will grow to the level:
\begin{widetext}
\begin{equation}\label{Nk}
  |b_k|^2 = |b_{-k}|^2 = N_k \equiv 
  \frac{
    \sqrt{(\gamma h)^2\xi_k^2N_0 - \left \{ (\omega_k-\omega_s) \eta_0 + (\omega_0-\omega_s)  \eta_k \right \} ^2} 
    + (\omega_0-\omega_s)(\omega_k-\omega_s) -\eta_0 \eta_k}
  {N_0 \xi_k^2}
  \,,\end{equation}
\end{widetext}
that provide the effective damping to fix $N_0$ at the level given by
Eq.~(\ref{Bth}).  The number of magnons in the majority mode $N_k$
changes {\em abruptly} with the change of the excited mode due to the
sweep of the magnetic field.  In the differential graphs this will
lead to a singularity at the points, where the transfer from one
excited mode to another takes place. In reality, due to the presence
of the thermal noise and other nonlinear processes (that are ignored
here) one will observe a sharp (intrinsically broaden) peak at the
points where the transfer from one excited mode to another occurs.

To see this more clearly, let us assume that coupling coefficients of
all modes are the same, $\xi_k = \xi$. Then
the changes of the parametrically excited modes (say, modes 1 and 2)
will be at such a magnetic field, that $(\omega_1-\omega_s) = -(\omega_2-\omega_s) = \Delta\omega/2$
(see Eq.~(\ref{Bth}).

Assuming that $|\omega_0-\omega_s| \ll \eta_k$ and $\Delta\omega \ll \eta_k$ we can write $N_0 \approx \eta_k/\xi$ and,
approximately,
\begin{equation}
  N_{1,2} \approx (\zeta - 1)\frac{\eta_0}{\xi}
  \pm \frac{(\zeta - 1)}{\zeta}\frac{(\omega_0-\omega_s) \Delta\omega}{2\xi\eta_k}
  \,,\end{equation}
where $\zeta \equiv h/h_{\rm cr}$ is the supercriticality parameter of the microwave magnetic field.

Thus, the total number of spin wave modes differ by
\begin{equation}
  \Delta N_t \equiv N_1 - N_2 = \frac{(\zeta - 1)}{\zeta} \frac{(\omega_0-\omega_s) \Delta\omega }{\xi\eta_k}
  \,,\end{equation}
which will give small "steps" (of the order of 1\% of $N_t$) in the behavior of $M_z(H_\text{ext})$ at different "integral" characteristics and ultranarrow peaks at "differential" characteristics of $M_z$ vs. $H_\text{ext}$. These analytical findings are in good agreement witht the experimental observed in Fig.\ref{fig1}.

In conclusion, we have achieved FMR spectroscopy of parametric
magnons.  The observed linewidth of 30~mG corresponds to a lifetime of
the order of 2$\mu$s, which should be compared with the spin-lattice
relaxation rate $T_1 = 106$ns usually observed in the lineare regime
\cite{klein:03}.  Such long lifetime starts to become comparable to
the one found in nuclear magnetic resonance using shimmed magnets.
This result opens up the possibility of making ultra-stables Yig Tuned
Oscillators or ultra-sensitive absolute magnetic field detectors using
these parametric resonances.


\end{document}